\documentclass[aps,prl,preprint,superscriptaddress]{revtex4}
\usepackage{graphicx}
\usepackage{color}

\begin{document}

\title{Direct view on the ultrafast carrier dynamics in graphene}
\author{Jens Christian Johannsen$^{\ast}$}
\affiliation{Institute of Condensed Matter Physics, \'Ecole Polytechnique F\'ed\'erale de Lausanne (EPFL), Switzerland}
\author{S\o ren Ulstrup$^{\ast}$}
\affiliation{Department of Physics and Astronomy, Interdisciplinary Nanoscience Center (iNANO), Aarhus University, Denmark}
\author{Federico Cilento}
\author{Alberto Crepaldi}
\affiliation{Sincrotrone Trieste, Trieste, Italy}
\author{Michele Zacchigna}
\affiliation{IOM-CNR Laboratorio TASC, Area Science Park,Trieste, Italy}
\author{Cephise Cacho}
\author{I. C. Edmond Turcu}
\affiliation{Central Laser Facility, STFC Rutherford Appleton Laboratory, Harwell, United Kingdom}
 \author{Emma Springate}
\affiliation{Central Laser Facility, STFC Rutherford Appleton Laboratory, Harwell, United Kingdom}
\author{Felix Fromm}
 \affiliation{Lehrstuhl f{\"u}r Technische Physik, Universit{\"a}t Erlangen-N{\"u}rnberg,Germany}
  \author{Christian Raidel}
 \affiliation{Lehrstuhl f{\"u}r Technische Physik, Universit{\"a}t Erlangen-N{\"u}rnberg,Germany}
 \author{Thomas Seyller}
 \affiliation{Lehrstuhl f{\"u}r Technische Physik, Universit{\"a}t Erlangen-N{\"u}rnberg,Germany}
\author{Fulvio Parmigiani}
\affiliation{Sincrotrone Trieste, Trieste, Italy}
\affiliation{Department of Physics, University of Trieste, Italy}
\author{Marco Grioni}
\affiliation{Institute of Condensed Matter Physics, \'Ecole Polytechnique F\'ed\'erale de Lausanne (EPFL), Switzerland}
\author{Philip Hofmann}
\affiliation{Department of Physics and Astronomy, Interdisciplinary Nanoscience Center (iNANO), Aarhus University, Denmark}
\affiliation{email: philip@phys.au.dk \\ $^{\ast}$ These authors contributed equally to the work.}

\date{\today}
 \begin{abstract}
The ultrafast dynamics of excited carriers in graphene is closely linked to the Dirac spectrum and plays a central role for many electronic and optoelectronic applications. Harvesting energy from excited electron-hole pairs, for instance, is only possible if these pairs can be separated before they lose energy to vibrations, merely heating the lattice. While the hot carrier dynamics in graphene could so far only be accessed indirectly, we here present a direct time-resolved view on the Dirac cone by angle-resolved photoemission (ARPES). This allows us to show the quasi-instant thermalisation of the electron gas to a temperature of more than 2000~K; to determine the time-resolved carrier density; to disentangle the subsequent decay into excitations of optical phonons and acoustic phonons (directly and via supercollisions); and to show how the presence of the hot carrier distribution affects the lifetime of the states far below the Fermi energy.
 \end{abstract}

\maketitle
\begin{figure}
\includegraphics[width=.99\textwidth]{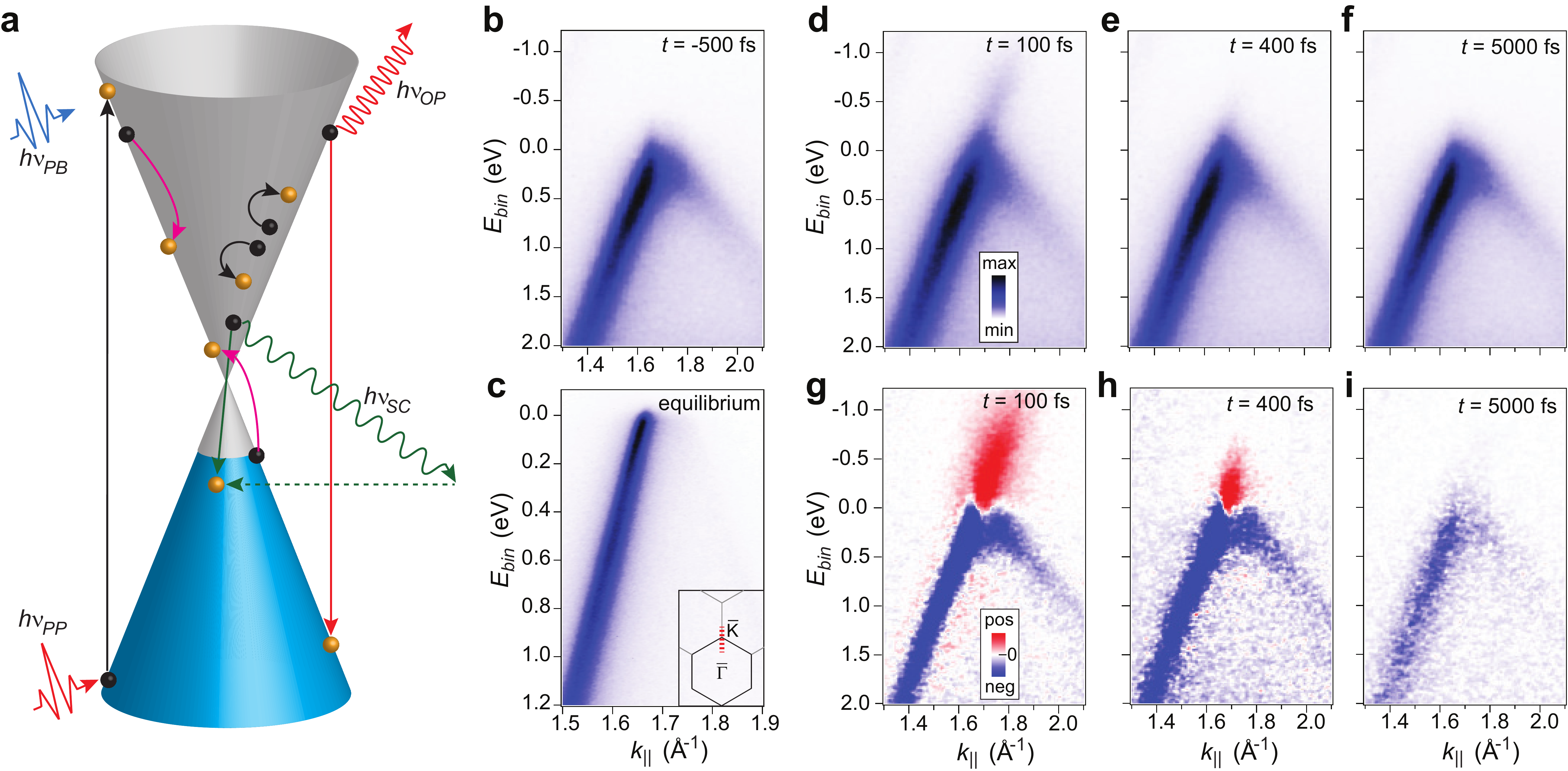}\\
\caption{\textbf{a} Dirac spectrum of graphene with an optical excitation $h\nu_{PP}$ and a probe $h\nu_{PB}$. Following the excitation, electrons (black spheres) thermalise via impact scattering (magenta curved arrows) or Auger recombination (black  curved arrows). Decay into holes (orange spheres) in the valence band can be mediated by emission of optical phonons (red wiggled line) with energy $h\nu_{OP}$ or supercollisions involving acoustic phonons (green wiggled line) with energy $h\nu_{SC}$ recoiling on impurities (dashed green line). \textbf{b} Photoemission intensity around the Dirac cone for a negative time delay (before the pump pulse). \textbf{c} Equilibrium high-resolution spectrum from the same sample taken with synchrotron radiation. Insert: Brillouin zone with measurement direction (dashed red line). \textbf{d} -\textbf{f} Spectra taken at different time delays. \textbf{g} - \textbf{i} Corresponding difference spectra, i.e. change with respect to the spectrum before the pump pulse. }
  \label{fig:1}
\end{figure}

Exploiting the unique properties of graphene in electronic or optoelectronic devices inevitably involves the generation of hot carriers, i.e. electrons or holes with high energies compared to the Fermi energy \cite{Castro-Neto:2009,Bonaccorso:2010,Gabor:2011}.  Such hot carriers are thought to thermalise via the electron-electron interaction on a time-scale of $\approx$ 30~fs \cite{Kampfrath:2005,Breusing:2009,Lui:2010}, leaving the electronic system at a well-defined temperature that can be substantially above the lattice temperature. If an initial electron-hole pair had been created by photo absorption, the thermalisation can be accompanied by carrier multiplication, i.e. the generation of more than one electron-hole pair per absorbed photon \cite{Winzer:2010,Song:2011d}. The hot electronic system can then efficiently lose energy by the fast emission of optical phonons, but this channel is blocked for electrons with too low energies ($\lesssim 200$~meV) or when the temperature of the optical phonons reaches that of the electrons \cite{Butscher:2007,Tse:2009,Wang:2010a,Tielrooij:2012}. Then a slow decay via acoustic phonons sets in. Indeed, this decay is so slow and inefficient that three-body collisions with defects and acoustic phonons, so-called supercollisions, can become the dominant contribution rather than the exception \cite{Song:2012e,Betz:2012,Graham:2012}. This hot carrier dynamics, summarised in Fig. \ref{fig:1}\textbf{a}, has been studied intensely but so far the ultrafast decay could only be probed indirectly, for example through the change of the transmissivity in a pump-probe experiment. 

Time and angle-resolved photoemission spectroscopy (TR-ARPES) offers an unparalleled access to the detailed carrier dynamics in the spectral function of solids and has been successfully applied to study many systems (see for example Refs. \cite{Perfetti:2007,Rohwer:2011}). It has the potential to directly provide the out-of-equilibrium single-particle spectral function, the statistical distribution of the carriers and its time evolution. However, for graphene the application of this technique is hampered by the position of the Dirac cone at the $\bar{K}$ point, far from the centre of the Brillouin zone. In order to collect TR-ARPES data from the Dirac cone, the final state electrons have to have a sufficiently high $k_{\parallel}$ and this can only be achieved for photon energies higher than $\approx 16$~eV, significantly higher than commonly available in conventional ultrafast laser sources \cite{Rohwer:2011,Stange:2013}. In order to overcome this hurdle, we have used the coherent high harmonics generated by laser light at the Artemis facility at the Central Laser Facility / Rutherford Appleton Laboratory. Hot electrons in graphene are generated by a 30~fs pulse with a photon energy of $h\nu=$0.95~eV.  This pump pulse leads to a transition from the valence band to the conduction band of graphene, as illustrated in Fig. \ref{fig:1}\textbf{a}. The spectral function around the Dirac cone is then measured by another laser pulse of $h\nu=$ 33.2~eV, following the first pulse by a variable time delay.

In order to closely approach the electronic properties of pristine, free-standing graphene, we use a sample of hydrogen-intercalated so-called quasi free-standing monolayer graphene (QFMLG) on SiC \cite{Riedl:2009,Speck:2011}. The sample is slightly hole doped with a carrier concentration of $5\times 10^{12}$~cm$^{-2}$, placing the Dirac point 240~meV above the Fermi energy $E_F$.  QFMLG is characterised by high crystalline quality, sharp photoemission lines, weak electron-phonon coupling  \cite{Johannsen:2013} and an efficient decoupling from the substrate, such that even subtle many-body effects can be observed \cite{Bostwick:2010}. ARPES data from the Dirac cone of QFMLG are shown in Fig. \ref{fig:1}\textbf{b} for a negative time delay, i.e. before the excitation with the pump pulse. As a comparison, high-resolution ARPES measurements from the same sample are shown in Fig. \ref{fig:1}\textbf{c} \cite{Johannsen:2013}.

Fig. \ref{fig:1}\textbf{d}-\textbf{f} show TR-ARPES spectra taken at different delays after the pump pulse (see also Supplementary Movie S1) and Fig. \ref{fig:1}\textbf{g}-\textbf{i} the corresponding difference between these spectra and the spectrum before the arrival of the pump pulse.  A clear increase (decrease) of the spectral intensity in the conduction band (valence band) is already discernible in the raw data and even clearer in the corresponding difference. Note that the strong intensity asymmetry between the two branches of the Dirac cone in every spectrum is a well-known interference phenomenon in ARPES from graphene \cite{Shirley:1995b}. 

Immediately after the pump pulse, one should observe a depletion (increase) of spectral intensity around the energy of the initial (final) states of the excitation, or at least a separate distribution for electrons and holes \cite{Breusing:2009} but it has been shown that the electron-electron interaction leads to a thermalisation of the carriers on a time scale of $\approx 30$~fs, faster than the experimental time resolution \cite{Kampfrath:2005,Breusing:2009,Lui:2010}. If this is so, we should only ever observe an electron gas with a thermal Fermi-Dirac (FD) distribution.

\begin{figure}
 \includegraphics[width=.99\textwidth]{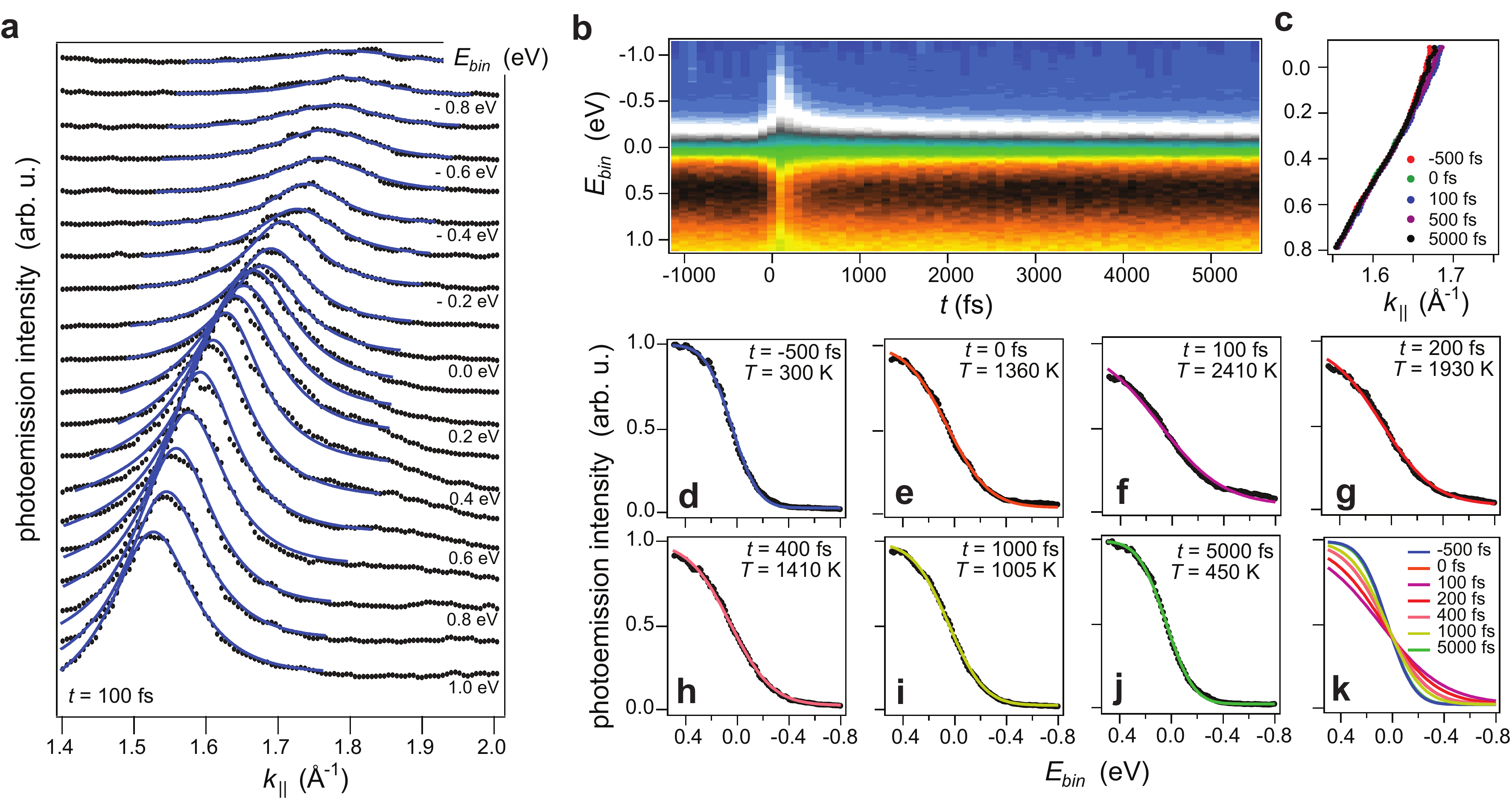}\\
\caption{\textbf{a} Spectrum from Fig. \ref{fig:1}\textbf{d} shown as momentum distribution cuts (only a sub-set is shown) and fits of the individual cuts by Lorentzians to determine intensity and position. \textbf{b} Intensity obtained in this way plotted as a function of energy and time delay. \textbf{c} Fitted MDC peak positions at various time delays. \textbf{d} - \textbf{j} Cuts through the data in \textbf{b} with fits by Fermi-Dirac distributions convoluted with a Gaussian to account for the energy resolution. \textbf{k} Fitted distributions at selected time delays.}
  \label{fig:2}
\end{figure}

We have direct access to the distribution of the carriers using the procedure introduced in Fig. \ref{fig:2}. For each delay time, we slice the image into momentum distribution curves (MDCs), i.e. the photoemission intensity at a given binding energy as a function of $k_{\parallel}$. A stack of such momentum distribution curves for a time delay of 100~fs can be seen in Fig. \ref{fig:2}\textbf{a}. Each MDC is then fitted with a Lorentzian peak, giving the peak position and the integrated intensity of the peak as a function of binding energy. The MDC intensity as a function of energy and time is shown in Fig. \ref{fig:2}\textbf{b}. It can be viewed as the statistical distribution of the carriers, tracking the dispersion of the state. The analysis of the MDC peak position confirms that there is no time-dependent change of the band structure, as expected for the low fluence employed here. Note that the apparent upward bending of the dispersion near $E_F$ is most likely caused by the finite energy resolution \cite{Kirkegaard:2005}.

The direct access to the electron distribution confirms the ultrafast thermalisation of the carriers. The data in Fig. \ref{fig:2}\textbf{b} are well described by a FD distribution for all time delays, as illustrated by a few cuts in Fig. \ref{fig:2}\textbf{d}-\textbf{k}. This confirms that the thermalisation of the carriers happens at a time scale that is shorter than the experimental time resolution, permitting the direct determination of an electronic temperature $T_e(t)$. Interestingly, deviations from the FD behaviour can be observed immediately after the main pump excitation at time delays between 0 and 200~fs where the photoemission intensity far above $E_F$ is higher than expected for the FD distribution. While this is a strong indication of an out-of-equilibrium situation before thermalisation, the limited time resolution does not permit its more detailed investigation.  

\begin{figure}
\includegraphics[width=.55\textwidth]{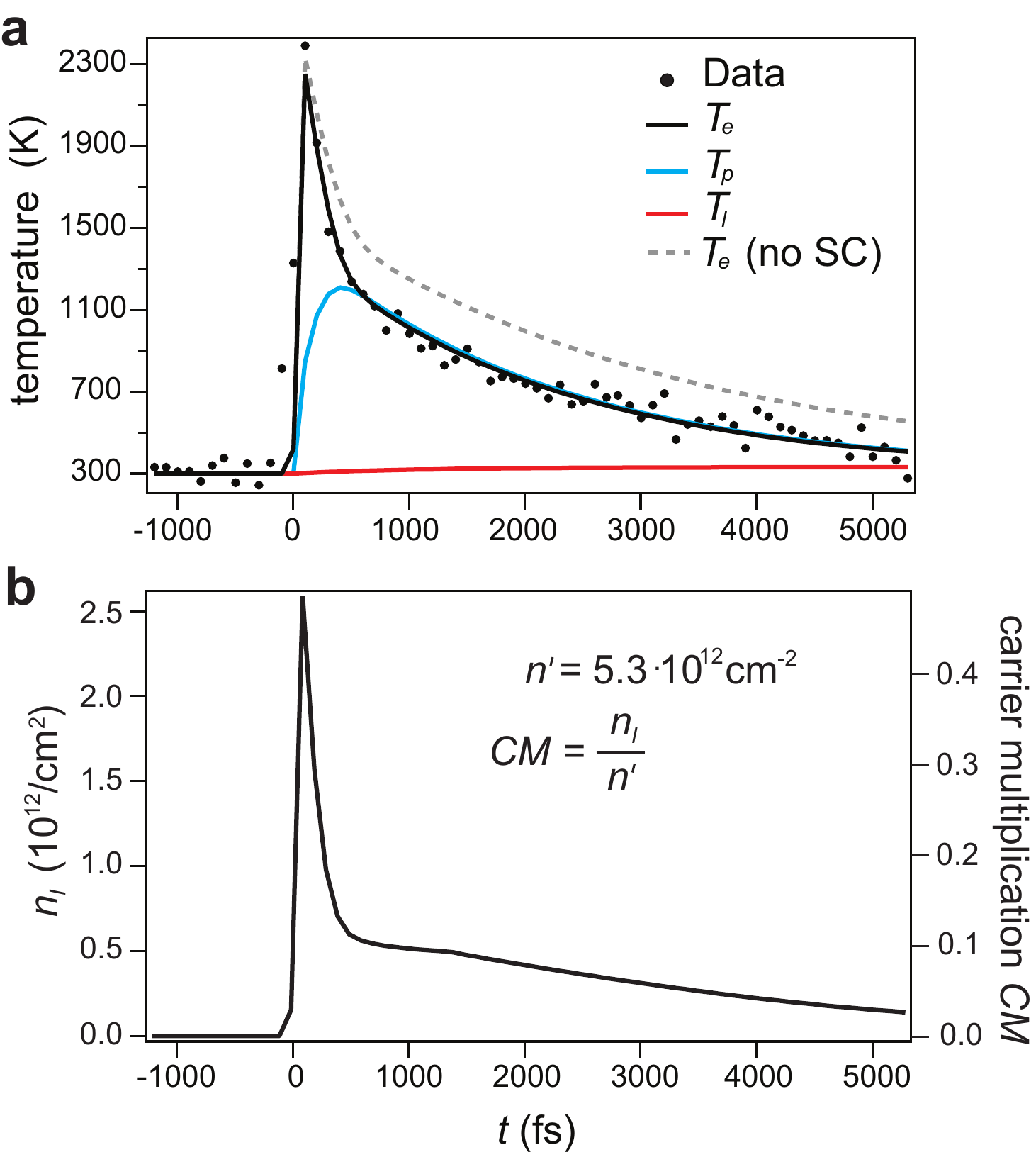}\\
\caption{\textbf{a} Electronic temperature $T_e(t)$ as a function of time delay (markers). Fitting the data to the three temperature model provides an electronic temperature fit with (black line) and without (dashed grey line) the supercollision (SC) term. From the fit we also obtain the temperatures of the optical phonons (blue) and the acoustic phonons (red). \textbf{b} Photo-induced carrier density $n_I$ and carrier multiplication calculated from the fitted electronic temperature.}
  \label{fig:3}
\end{figure}

The electronic temperature extracted from this procedure is shown in Fig. \ref{fig:3}\textbf{a}. The most simple analysis shows that the data cannot be fitted by one single exponential decay but by a double exponential decay involving two time constants, $\tau_1 \approx 160$~fs and $\tau_2 \approx 2700$~fs. These are tentatively assigned to processes involving optical phonons and acoustic phonons / supercollisions, respectively. In order to analyse this further, we  employ a phenomenological three temperature model in a similar form as recently applied to a TR-ARPES study of a high-T$_{C}$ superconductor \cite{Perfetti:2007,DalConte:2012}. The model assumes that three different temperatures exist in the system: the electronic temperature $T_e$, the temperature of the optical phonons $T_p$ and that of the acoustic phonons $T_l$. This leads to three coupled differential equations:
\newpage
\begin{eqnarray}
\frac{\partial T_e}{\partial t} &=& + \frac{S(t)}{\beta} - \frac{\pi\lambda_1g(\mu(T_e)) \Omega_{Ein}^3}{\hbar}\frac{n_e - n_p}{C_e} - 9.62\frac{\lambda_2g(\mu(T_e))k_B^3}{\hbar k_F l}\frac{T_e^3-T_l^3}{C_e} \nonumber \\ && - \pi\lambda_2\hbar g(\mu(T_e)) k_F^2 v_s^2 k_B\frac{T_e - T_l}{C_e}, \label{eqn:1}
\end{eqnarray}
\begin{equation}
\frac{\partial T_p}{\partial t} = \frac{C_e}{C_p} \frac{\pi\lambda_1g(\mu(T_e)) \Omega_{Ein}^3}{\hbar}\frac{n_e - n_p}{C_e} - \frac{T_p - T_l}{\tau_p},  \label{eqn:2}
\end{equation}
\begin{equation}
\frac{\partial T_l}{\partial t} = \frac{C_e}{C_l} \left(9.62\frac{\lambda_2g(\mu(T_e))k_B^3}{\hbar k_F l}\frac{T_e^3-T_l^3}{C_e} + \pi\lambda_2\hbar g(\mu(T_e)) k_F^2 v_s^2 k_B\frac{T_e - T_l}{C_e} \right) + \frac{C_p}{C_l}\frac{T_p - T_l}{\tau_p},  \label{eqn:3}
\end{equation}
where $C_e$, $C_p$ and $C_l$ are the heat capacities, $\lambda_1$ and $\lambda_2$ are the electron-phonon coupling constants to the optical and the acoustic phonons, respectively, $\Omega_{Ein}$ is the energy of the optical phonons, modelled as Einstein oscillators, $n_e$ and $n_p$ are distribution functions given as $(e^{\Omega_{Ein}/k_BT_(e/p)}-1)^{-1}$, $g(\mu(T_e))$ is the density of states evaluated at the temperature dependent chemical potential $\mu(T_e)$, $l$ is the mean free path of the electrons, $v_s$ is the sound velocity in graphene and $k_F$ is the Fermi wave vector with respect to the $\bar{K}$ point.

Details of the model are given in the Supplementary Material. Here we concentrate on the general form of the equations. The first term in (\ref{eqn:1}) stems from the heating by the pump pulse $S(t)$ that has a Gaussian distribution with a full width at half maximum (FWHM) of 30~fs. $\beta$ is a constant with the dimension of a heat capacity but note that the heat capacity of the electronic system is not properly defined during the pump phase. After the excitation, the thermalisation of the electrons is assumed to be sufficiently fast that a definition of $T_e$ is meaningful and $C_e$ can be defined. The electrons can now lose energy through either coupling to optical phonons (second term in (\ref{eqn:1})) or to acoustic phonons, either via supercollisions \cite{Song:2012e} or directly \cite{Bistritzer:2009} (remaining two terms). The corresponding equations for $T_p$ and $T_l$ describe how these sub-systems are heated up by the energy that is lost in the electronic system. In addition to this, it is also possible for optical phonons to undergo anharmonic decay into acoustic phonons with a characteristic lifetime $\tau_p$.

The unknown parameters in the equation system are the coupling constants $\lambda_{1,2}$ and the scaling parameter $\beta$. By adjusting these, it is possible to obtain an excellent fit to the data (black line in Fig. \ref{fig:3}\textbf{a}). Note that $\beta$ only determines the initial temperature, it has no influence on the curve after the pump pulse duration. We find that the coupling constants are very small with $\lambda_1 = 0.052$ and $\lambda_2 = 0.0042$, as expected for lightly doped graphene \cite{Calandra:2007,Ulstrup:2012}. Indeed, our value for $\lambda_1$ is consistent with the electron-phonon coupling strength determined in Ref. \cite{Tielrooij:2012} and so is the fact that most of the incoming energy is absorbed by the electronic system.

We can also test if it is possible to fit the curve in the absence of supercollision processes by removing the corresponding term. This results in the grey dashed line. It clearly shows that the supercollisions are essential once $T_p = T_e$, the point at which cooling by optical phonons is exhausted. We also observe that the lattice temperature increases very little, which can be explained by its much higher heat capacity, allowing it to act like a perfect heat bath. 

Given the fact that the system follows a FD distribution with a well-defined temperature, we can calculate the time-dependent photo-induced carrier density $n_I$, i.e. the number of photo-excited electrons above $E_F$ (see Fig. \ref{fig:3}\textbf{b} and details in the Supplementary Material). From $n_I$ and a calculation of the absorbed energy at our maximum electronic temperature we arrive at an estimate of the number of electron-hole pairs  $n'$ generated by the pump pulse. This allows us to quantify the carrier multiplication ($CM$) in the system, i.e. the number of electron hole pairs generated for every photon absorbed. In our case, $CM$ remains below unity, merely reaching $\approx 0.5$, consistent with theoretical predictions \cite{Winzer:2012}. Note that these predictions also show that $CM$ can reach values well above unity for the lower fluence values and higher photon energies actually relevant for photoelectric applications. Here, because of the need for a minimum fluence for a pump-probe experiment, the CM $ > 1$ regime is outside our experimental conditions.

\begin{figure}
 \includegraphics[width=.55\textwidth]{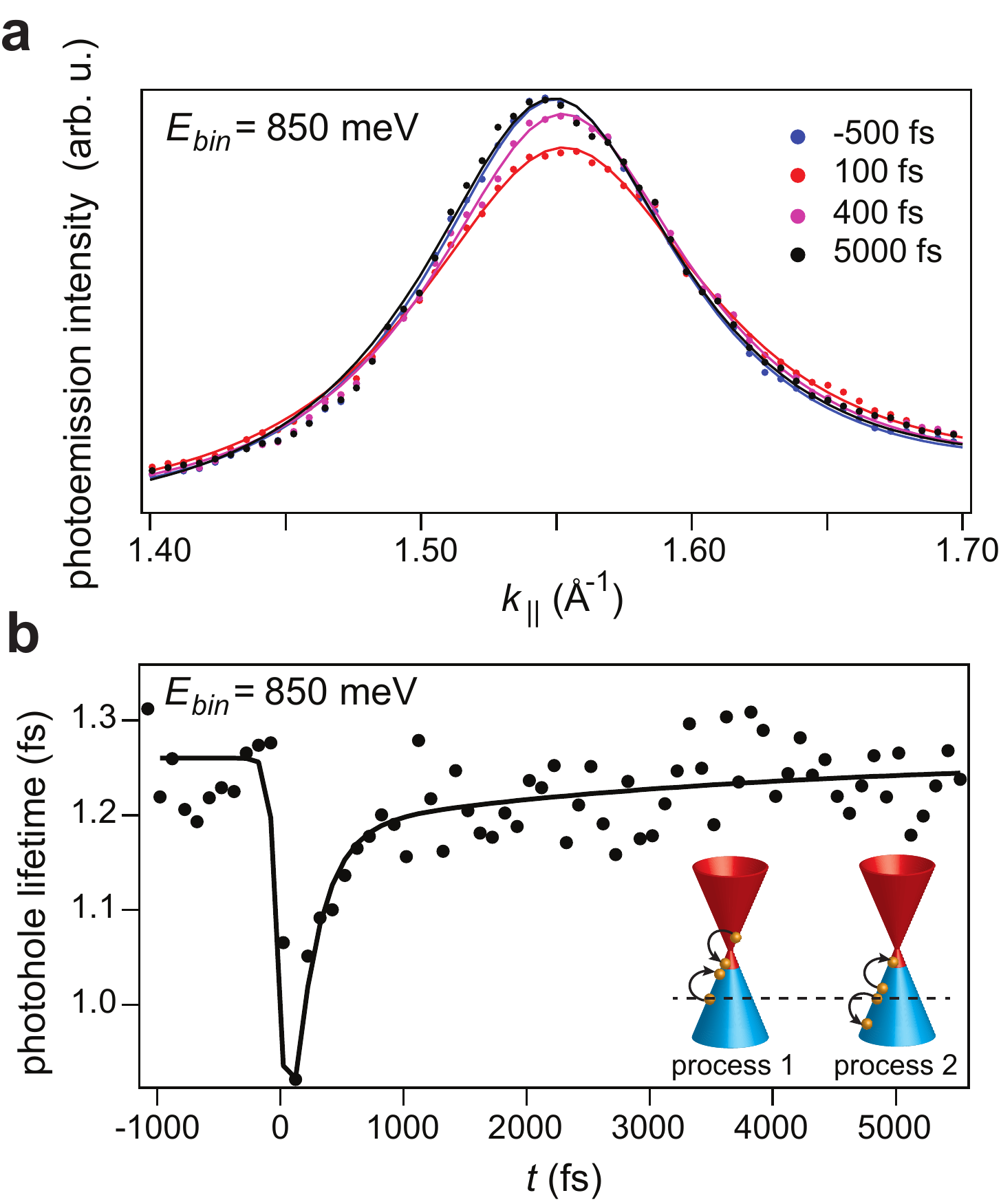}\\
\caption{\textbf{a} MDC cuts through the spectral function at a binding energy of 850~meV for different time delays. \textbf{b} Lifetime of the hole state as a function of time. The line is a fit to an exponential model with a time constant of 200~fs. Inserts: Dirac spectrum after excitation with hot electrons above and excess holes below the doping level. This situation gives rise to two hole-hole recombination processes (arrows). The dashed line marks the energy of the photohole.}
  \label{fig:4}
\end{figure}

The linewidth of states in ARPES can give detailed information about the many-body effects in the hole spectral function. In particular, the MDC width can be related to the imaginary part of the self-energy or, equivalently, to the lifetime of the photohole. We can therefore not only investigate the ultrafast redistribution of carriers but also the effect this has on the energy and momentum-resolved lifetime of a given state. Fig. \ref{fig:4}\textbf{a} shows a cut through the spectral function deep in the occupied states, at a binding energy of 850~meV, for different time delays. After the pump pulse, not only does the total intensity decrease due to the lower value of the FD distribution, we also observe an increase of the MDC width from $\approx$ 0.115~\AA$^{-1}$ to  $\approx$ 0.141~\AA$^{-1}$. This change in width corresponds to the opening of an additional decay channel for the photohole. We estimate the lifetime from the MDC width and find that the initial lifetime $\tau_0 \approx 1.26$~fs is reduced to $\tau_1 \approx 0.92$~fs. The characteristic lifetime for the new decay channel $\tau_d$ can be estimated from $1/\tau_1 = 1/\tau_0 + 1/\tau_d$ to be $\tau_d \approx$ 3.41~fs. Fig. \ref{fig:4}\textbf{b} shows the time-dependence of the photohole lifetime estimated from the MDC width, showing that the new decay channel of the photohole closes very quickly on a timescale of 200~fs. This observation permits us to narrow down the origin of the decay channel. Consider the possible hole-hole scattering processes shown in Fig. \ref{fig:4}\textbf{b} \cite{Rana:2007}. Process 1 involves scattering with holes at a lower binding energy than the photohole while process 2 requires holes with a higher binding energy. Since the number of these high energy holes decreases very rapidly \cite{Crepaldi:2012}, the time-dependence of the MDC broadening suggests that the new decay channel is dominated by process 2-like scattering events. Note that the decay in  Fig. \ref{fig:4}\textbf{b} appears very similar to the initial decay of $T_e$, a process ascribed to the excitation of optical phonons. However, optical phonons are not directly responsible for the additional decay channel of the photohole because such processes are always possible; they are not enhanced by the presence of hot holes. The reason for the time scale being similar is that process 2 requires high energy holes and these are rapidly depleted during the initial cooling.

Concluding, we have been able to directly measure the time, energy and momentum-resolved statistical distribution of hot electrons in quasi free-standing graphene after a photo-excitation process. We confirm the ultrafast redistribution of carriers to a hot thermalised Fermi-Dirac function with no separate electron and hole distributions discernible within our time resolution. The decay process of this hot electron population proceeds with the expected involvement of optical and acoustic phonon excitations and we are directly able to reveal the role of supercollisions in this process. We quantify the number of induced electron-hole pairs and find that it stays under the threshold of carrier multiplication. This result is found to be consistent with quantitative predictions of carrier multiplication in graphene and suggests that this effect is likely to play a role in the low-fluence situation of actual devices. Finally, we show that TR-ARPES can disentangle the photohole lifetime, described by the single-particle spectral function, from the lifetime of the excited charge carrier population distribution.

\section{Methods}

TR-ARPES experiments were performed at the Artemis facility at the Central Laser Facility / Rutherford Appleton Laboratory \cite{turcu:2009}.Ê A 1~kHz Ti:sapphire amplified laser system provided ultrafast (30 fs FWHM) infrared pulses at 785 nm, with an energy per pulse of 10~mJ. For angle-resolved photoemission, 20 \% of the laser energy was used to generate high-order harmonic femtosecond XUV pulses in a pulsed jet of argon gas. The time-preserving monochromator \cite{Frassetto:2011} selected the 21st harmonic with a photon energy of 33.2~eV. The remaining power was utilised to drive an optical parametric amplifier (HE-Topas), which provided tuneable laser pump pulses (30~fs). We chose a pump energy of 0.95 eV (1300 nm). The pump fluence was approximately 346~$\mu$J/cm$^2$, and the beam was polarised perpendicular to the scattering plane (s-polarised) in order to avoid laser-assisted photoemission effects. For further details of the experimental set-up see Refs. \cite{Petersen:2011,Frassetto:2011}. Time, energy and angular resolution were set to 60 fs, 350 meV and 0.3$^{\circ}$, respectively. The high-resolution static ARPES measurements in Fig. \ref{fig:1}\textbf{c} were measured on the SGM-3 beamline of the synchrotron radiation source ASTRID in Aarhus \cite{Hoffmann:2004}.

H-intercalated epitaxial graphene on SiC(0001) was prepared \emph{ex-situ} by the methodology given in Ref. \cite{Speck:2010}, and was cleaned by annealing to 550~K in ultra-high vacuum in order to remove absorbed water. The sample was held at room temperature throughout the entire experiment.

We note that the extraction of the carrier distribution used in Fig. \ref{fig:2} is somewhat cumbersome compared to the frequently used method of  analysing a cut through the photoemission intensity at a single angle or $k_{\parallel}$. However, in the present case this simpler method leads to very inaccurate results for the electronic temperature, underestimating it by up to 1000~K.

For the three temperature model in equations (\ref{eqn:1})-(\ref{eqn:3}), the following parameters were used: The optical phonons were approximated by Einstein oscillators with $\Omega_{Ein}=200$~meV, and the anharmonic decay was set at $\tau_p = 2.5$~ps (see Ref. \cite{Wang:2010a}). The mean free path of the electrons $l$ was determined by high-resolution ARPES measurements from the MDC full width at half maximum at $E_F$, which is given by $1/ l$. 

The lifetimes $\tau$ in connection with Fig. \ref{fig:4} were also determined from the MDCs width and the slope of the band $v=dE(k)/dk$, using that $l=v \tau$. In the actual determination of $\tau$, however, the finite momentum and energy resolution have also been taken into account.


\section{Acknowledgements}
We thank Lars Bojer Madsen, Torben Winzer and Ermin Mali\'c for helpful discussions, and Phil Rice, Natercia Rodrigues and Richard Chapman for technical support during the beamtime at the Artemis Facility. We gratefully acknowledge financial support from the VILLUM foundation, The Danish Council for Independent Research / Technology and Production Sciences. Access to the Artemis Facility was funded by Laserlab Europe. Work in Erlangen was supported by the European Union through the project ConceptGraphene, and by the German Research Foundation in the framework of the SPP 1459 ÔGrapheneÔ. F.P., F.C., A.C and M.Z. acknowledge Barbara Ressel for the technical support and discussions and the Italian Ministry of University and Research for providing the fundings under the Grants No. FIRBRBAP045JF2 and No. FIRB-RBAP06AWK3. 




\end{document}